\documentclass[twocolumn,aps,showpacs,amsmath,amssymb]{revtex4}
\usepackage{bm}
\usepackage{graphicx}

\begin{document}

\newcommand{\hmin}{h_{\mathrm {min}}}
\newcommand{\hfront}{h_{\mathrm {front}}}
\newcommand{\hside}{h_{\mathrm {side}}}
\newcommand{\rn}{R_{\mathrm {N}}}

\title{Breakup of Air Bubbles in Water: \\ Memory and Breakdown of Cylindrical Symmetry}

\author{Nathan~C.~Keim}
\email{nkeim@uchicago.edu}
\author{Peder~M{\o}ller}
\author{Wendy~W.~Zhang}
\author{Sidney~R.~Nagel}
\affiliation{James Franck Institute, University of Chicago, 929 E.\ 57th St., Chicago, IL 60637, USA}

\date{\today}

\begin{abstract}

Using high-speed video, we have studied air bubbles detaching from an
underwater nozzle. As a bubble distorts, it forms a thin neck which
develops a singular shape as it pinches off. As in other singularities, the
minimum neck radius scales with the time until breakup. However, because the
air-water interfacial tension does not drive breakup, even small
initial cylindrical asymmetries are preserved throughout the collapse. This
novel, non-universal singularity retains a memory of the nozzle shape, size and
tilt angle. In the last stages, the air appears to tear instead of pinch. 

\end{abstract}


\maketitle

The delightful tingling felt when drinking carbonated beverages, the glee of
children blowing bubbles in a bathtub, and the importance of deep underwater
fissures venting gasses into the oceans hint at the richness and significance
of bubble formation in determining the texture and composition of our world.
However, the process by which a bubble is formed is still full of surprises. A
drop or bubble breaks up by forming a neck that thins to atomic dimensions, a
process described as an approach towards a singularity where physical
quantities such as stress or pressure grow infinitely large. Singularities
often organize the overall dynamical evolution of nonlinear systems. Each
symmetry in nature implies an underlying conservation law, so that the
symmetries of the singularity associated with pinch-off naturally have
important consequences for its dynamics. It was previously believed
\cite{eggers94,shi94,eggers97,chen97,day98,lister98,zhang99,cohen99,cohen01,chen02,leppinen03,sierou03}
that the pinching neck of any drop or bubble would become cylindrically
(\emph{i.e.}\ azimuthally) symmetric in the course of pinch-off. Recently,
pinching necks of air in water were observed to lose cylindrical symmetry in
the course of detachment \cite{keim05,bergmann06}. 

Here we show that this loss of symmetry is caused by a new form of memory in
singular dynamics: even a small asymmetry in the initial conditions is
preserved throughout bubble detachment. This novel singularity retains a memory
of the nozzle shape, size and tilt angle. The asymmetry can be made so great
that the air appears to tear. This symmetry breaking may be important in
numerous applications \cite{sutherland88,rubio02,theander04}, and for
understanding other physical processes which are modeled as the formation of a
singularity, such as star or black hole formation \cite{liebling96} and
supernova explosions \cite{plewa04}. Thus our experimental observation of the
breakdown of cylindrical symmetry in the air bubble demonstrates a new view of
dynamical singularities that may be relevant even on a celestial scale.    

\begin{figure}
\includegraphics{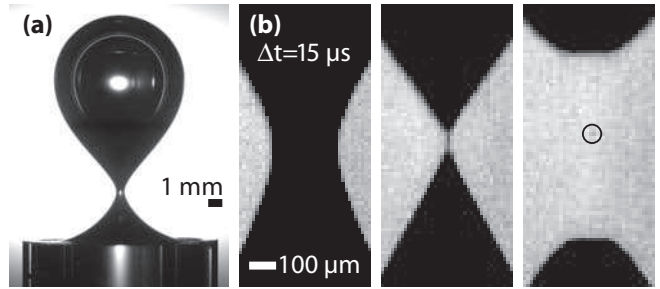}

\caption{Pinch-off of an air bubble from a level underwater, circular nozzle
with radius $\rn = 4.1$~mm. \textbf{(a)}~The bubble appears dark, except for
bright optical artifacts. \textbf{(b)}~A magnified sequence of pinch-off. In
the last frame, the single $\sim$5~$\mu$m satellite bubble is circled. $\Delta
t$ gives the time between frames.}

\label{fig:clean}
\end{figure}

Singularities govern the dynamics in many familiar break-up events, such as the
dispersal of oil drops into vinegar during the making of a salad dressing, or
the dripping of water from a leaky faucet.  For many fluid pairs --- for
example, one viscous fluid breaking in a surrounding fluid of high
viscosity \cite{zhang99,cohen99,sierou03} --- the shape and dynamics of the
pinching neck depend solely on the fluid parameters, as the breakup forgets its
initial conditions on approaching the singularity.  Such universal behavior,
where the dynamics of pinch-off are dominated solely by the singularity that
lies at its end, was until recently thought to be the only way in which a fluid
could break apart.  However, Doshi \emph{et al.}\ \cite{doshi03} discovered an
exceptional form of pinch-off when an inviscid fluid pinches off inside a
viscous one. Here the axial curvature of the neck is preserved and a change in
nozzle size is remembered throughout breakup.  That any memory of initial
conditions persists is surprising and raises new questions about the possible
types of dynamics near a singularity: during pinch-off, how much and what kind
of information can be remembered about initial conditions, and how does this
memory influence the permanent structures that appear after the singularity has
been formed?  

In this letter, we show that when an air bubble breaks off from a
submerged nozzle, not only does the pinching neck of air retain an indelible
imprint of its initial geometry, but the initial azimuthal asymmetry alters the
number and subsequent trajectories of satellite bubbles. Extreme asymmetries in
the initial shape of the air bubble result in a new, fully three-dimensional
mode of pinch-off, in which the air tears apart in successive sharp jerks,
instead of pinching at a point. 

Normally, surface tension can be relied upon to restore the cylindrical
symmetry of a pinching liquid neck even when the initial conditions are
asymmetric. However, as first pointed out by Longuet-Higgins \emph{et
al.}\ \cite{longuet-higgins91} and subsequently elaborated by experiments and
simulations \cite{oguz93,burton05,gordillo05,keim05,leppinen05,thoroddsen05,bergmann06},
the detachment of an air bubble from a nozzle is not a
collapse driven by surface tension, but rather is an implosion due to a
pressure difference between the hydrostatic pressure in the water and the bubble pressure,
$\Delta P = \Delta \rho g a$. Here $a$ is the linear size of the bubble:
\begin{equation}
\label{eq:radius}
a = {\biggl( \frac{\rn \sigma}{g \Delta \rho} \biggr) }^{1/3}
\end{equation}
where $\rn$ is the nozzle radius, $\sigma$ is the surface
tension, $\Delta \rho$ is the difference in the densities of the liquid and the
air, and $g$ is the gravitational acceleration \cite{burton05}. 
The relevant experiment to indicate if
surface tension plays a role in the asymptotic dynamics, is to measure the
radius of the neck of air at its narrowest part, $\hmin$, as a function of
$\tau$, the time left to the singularity.  If the implosion dynamics persist
until breakup, then 
\begin{equation}
\label{eq:hmin}
\hmin = \beta (a^3 g)^{1/4} \tau^{\alpha_h}
\end{equation}
where $\beta$ is a numeric prefactor and $\alpha_h = {1/2}$. 

Burton \emph{et al.}\ \cite{burton05} reported excellent experimental
agreement with this result.  We note here that surface tension sets the
initial size of the bubble but plays no role in the dynamics.  This therefore
sets bubble detachment apart from all other breakup situations studied so far
where one or both fluids are viscous, or where two inviscid fluids differ
little in density, as well as from the inverse case of water in air.  In those
cases, the breakup is driven by surface tension and $\alpha_h \geq 2/3$
\cite{eggers97,chen97,day98,lister98,cohen99,zhang99,cohen01,chen02,leppinen03,doshi03,sierou03}.
Here we observe the consequences of a different driving mechanism on the
breakup dynamics. 

To create bubbles, we use a syringe pump to release air from circular
underwater nozzles with radii $\rn$ from 1.5~mm to 4.1~mm.  The nozzle and
water tank rest on a precision two-axis tilting platform, which allows us to
break and restore cylindrical symmetry. In contrast to
Bergmann \emph{et al.} \cite{bergmann06}, our bubbles from circular nozzles are
produced quasi-statically (0.03~s$^{-1}$), and so neither the Froude 
nor Bond number is a control parameter. Oblong nozzles are also used to
introduce more extreme asymmetries.  Bubbles are back-lit, and 
photographed with a Vision Research ``Phantom'' Version~7 camera at rates up to
130,000 frames/s. For each video frame, a computer traces the neck profile and
obtains $\hmin$.

\begin{figure}
\includegraphics{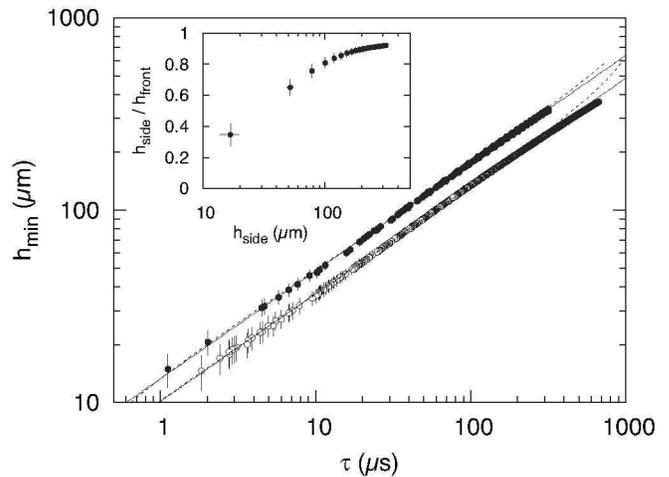}

\caption{Scaling of $\hmin$ versus $\tau$ for bubbles released in water from
circular nozzles of $\rn = 1.5$~mm (open circles) and 4.1~mm (closed circles).
Solid lines show power-law fits with exponent $\alpha_h = 0.56$, and dashed
lines show fits with logarithmic corrections \cite{gordillo05,bergmann06}.
Data are taken from multiple pinch-offs, and only data for $\tau
\leq$~230~$\mu$s are used for fitting. \textbf{Inset:} For the tilted-nozzle
image sequences shown in Fig.~\ref{fig:tilt}a, the radial size of the neck
scales differently in the ``front'' ($\hfront$) and ``side'' ($\hside$)
directions. }

\label{fig:hmin}
\end{figure}

Burton \emph{et al.}\ \cite{burton05} reported the remarkable observation that
instead of proceeding smoothly to zero radius, the neck of air abruptly breaks
apart in what they term a ``rupture'' at $\hmin \approx 25\ \mathrm{\mu m}$,
which they attribute to a Kelvin-Helmholtz instability that is intrinsic to the
dynamics.  However, as shown in Fig.~\ref{fig:clean}, when the nozzle is carefully
leveled, we find that the pinch-off appears to be cylindrically symmetric and
proceeds, without rupture, to scales below our camera resolution
($\sim$4~$\mu$m).  Figure~\ref{fig:hmin} shows $\hmin$ versus $\tau$ for two
nozzles. The data are well fit with a power law:
$\hmin \propto \tau^{\alpha_h}$, with $\alpha_h = 0.56 \pm 0.03$. This is
consistent with simulations of Leppinen \emph{et al.}\ who found $\alpha_h
\approx {0.55}$ \cite{leppinen05} but exclude $\alpha_h = 0.50$
\cite{longuet-higgins91,oguz93,burton05}. Our data cannot
distinguish pure power-law behavior with $\alpha_h = 0.56$ (solid lines in
Fig.~\ref{fig:hmin}) from a power law $\alpha_h = 0.50$ with logarithmic
corrections (dotted lines) that have been derived for Eqn.~\ref{eq:hmin}
\cite{gordillo05,bergmann06} --- a limitation also encountered by Bergmann
\emph{et al.}\ at low Froude numbers. However, these corrections are derived by
assuming a slender cylinder, which we believe is a poor approximation to our neck shape.
The pure power-law prefactor scales approximately as $\rn^{0.25}$, as predicted
by Eqns.~\ref{eq:radius} and~\ref{eq:hmin}. In a forthcoming paper we will
present a more detailed analysis of the dependence on $\rn$ and $\sigma$, and
the scaling of the entire neck profile in both the radial and axial
directions~\cite{keim-bubbles-unpub}. Here the exponent $\alpha_h$ is
considerably smaller than $2/3$. We suggest that the neck of an air bubble
could collapse so rapidly that the force due to surface tension would not keep
pace with the evolving dynamics; hydrostatic pressure and Bernoulli pressure
instead drive the breakup \cite{longuet-higgins91}\footnote{A comparison of
the Bernoulli pressure, which scales as $\rho \hmin^2/\tau^2 $ near breakup,
and Laplace pressure, which scales as $\sigma/\hmin$, shows that surface
tension effects are relevant in the final breakup dynamics if and only if
$\hmin$ decreases as $\tau^{2/3}$ or slower.}. We now show that, as break-up is
approached, surface tension is indeed no longer effective at erasing
cylindrical asymmetry and that this has profound effects. 

\begin{figure}
\includegraphics{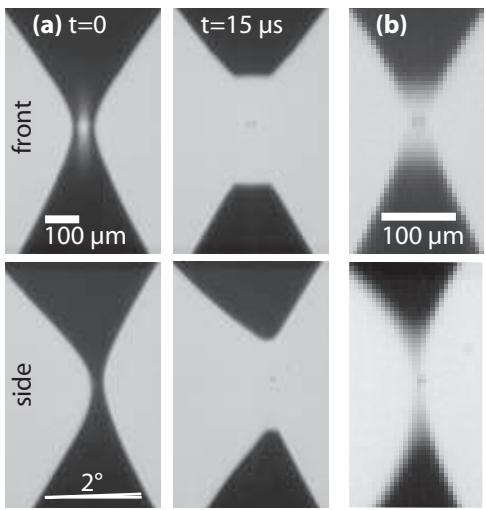}

\caption{``Front'' and ``side'' views of pinch-off from an $\rn = 4.1$~mm
nozzle tilted by 2$^\circ$. The nozzle's tilt is indicated in the bottom-left
frame; in the front images, it is tilted away from the camera.  Front and side
images are from separate sequences, but were selected to match length scales
closely. \textbf{(a)}~The neck is broadened before breakup, resembling a
crimped and bent double cone. A pair of satellite bubbles $\sim$15~$\mu$m in
diameter is produced. Afterward, the tip of each interface is bifurcated.
\textbf{(b)}~In these 2~$\mu$s exposures, the final form of the neck is
captured along with the initial positions of the satellite bubbles. The neck
reaches zero thickness while it is still $\sim$20~$\mu$m wide, resulting in a
what appears to be a ``rupture'' instead of a smooth, symmetric pinch-off.}

\label{fig:tilt}
\end{figure} 

We can test for a memory of asymmetry by tilting the nozzle slightly away from
the vertical axis. Figure~\ref{fig:tilt} represents pinch-off for tilt angles
$\agt 1^\circ$.  Even this small tilt flattens the neck sufficiently to
produce two satellite bubbles instead of the single one found with the leveled
nozzle (as shown in the last panel of Fig.~\ref{fig:clean}(b)); the interface
has an increasingly bifurcated appearance after pinch-off; and the breakup
event is no longer centered above the nozzle, but is shifted laterally away
from the direction of tilt (\emph{i.e.}\ rightward in the ``side'' views of
Fig.~\ref{fig:tilt}). For these pinch-offs, $\hmin$ also scales as a power
law, with $\alpha_h = 0.59 \pm 0.02$. Even deviations as small as $0.07^\circ$ from the
vertical axis give rise to visible lateral asymmetry in the interface profile
$\sim$100~$\mu$s after pinch-off. This smaller tilt corresponds to a
displacement of only 10~$\mu$m at the rim of the $\rn = 4.1$~mm nozzle used in
Fig.~\ref{fig:tilt}. The effects of tilting are most pronounced for the
larger nozzle.  The rupture observed by Burton \emph{et al.}\ \cite{burton05}
resembles our photographs taken with a small tilt (Fig.~\ref{fig:tilt}).  We
have seen that by carefully leveling the nozzle we can delay rupture to scales
below our resolution (4~$\mu$m); rupture is therefore unlikely to be caused by
an intrinsic Kelvin-Helmholtz instability at $\sim$$25\ \mathrm{\mu m}$ as
Burton \emph{et al.} suggested.

Tilting the nozzle also changes scaling, as shown in the inset
of Fig.~\ref{fig:hmin}. $\hmin$ is replaced by $\hfront$ and $\hside$,
corresponding to the views in Fig.~\ref{fig:tilt}. As $\hside$ becomes small,
the ratio $\hside / \hfront$ goes to zero. Above $\sim$10~$\mu$s before
breakup, scaling is the same as for a level nozzle, with $\lesssim$~10\%
difference in the power-law prefactor.

The satellite bubbles produced at breakup serve as tracers to indicate
subsequent liquid flow, as the buoyant rising of a 15~$\mu$m bubble
($\sim$$10^{-4}$~cm/s) is negligible compared to the observed velocities
($\sim$$80$~cm/s). Following pinch-off from a nozzle tilted by $2^\circ$,
satellite bubbles move upward and away from the direction of tilt (\emph{i.e.}\
rightward in Fig.~\ref{fig:tilt}). This motion suggests a cylindrical
asymmetry in fluid velocities around the breakup point: water that has traveled
farther to the breakup point is also moving faster. The pair of satellite
bubbles produced from a nozzle tilted by $2^\circ$ exhibit smaller-scale motion
in addition to the average motion just described: bubbles may circle about each
other laterally with a period $\sim$150~$\mu$s, indicating vertical vorticity.

\begin{figure}
\includegraphics{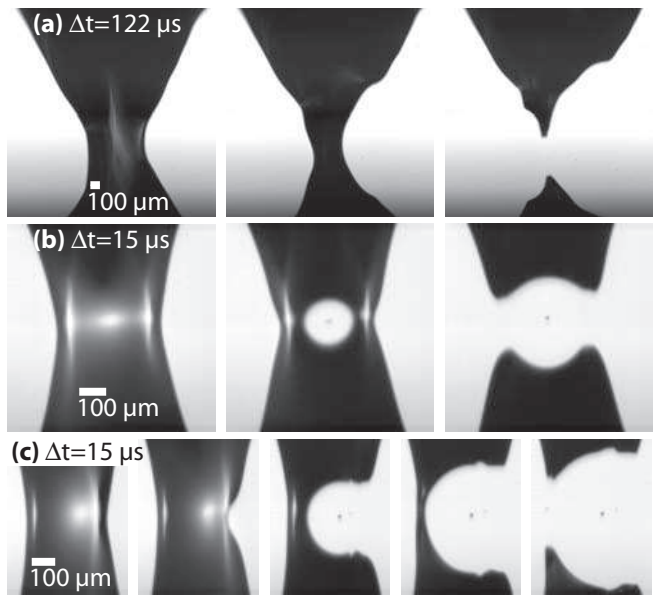}

\caption{Pinch-off of a burst of air from a slot-shaped nozzle, displaying
strong cylindrical asymmetry.  \textbf{(a)}~For slower bursts, the neck is
broadened with a scalloped profile.  In a series of discrete events, it
``tears'' iteratively from one or both sides.  \textbf{(b)}~For faster bursts,
the neck is more ribbon-like. The ribbon thins until coalescence is initiated,
creating a hole in the center of the neck. The remaining two columns of air
quickly break. Each of the topological transitions creates one or more
satellite bubbles. \textbf{(c)}~Under the conditions for (b), we also see
off-center tearing that combines aspects of (a) and (b).}

\label{fig:slot}
\end{figure} 

That pinch-off is sensitive to small deviations from cylindrical symmetry
suggests that gross asymmetry in initial conditions would lead to
correspondingly dramatic outcomes.   In addition, it could provide a visible
example of a fully three-dimensional breakup.   To test this, we used a nozzle
with an oblong opening, a 9.6~mm~$\times$ 1.6~mm slot with rounded ends. If
the timescale of bubble inflation is shorter than that for capillary waves
($\sim$100~ms), pinch-off begins asymmetrically. To achieve rapid
inflation (40~ms), we operate a valve to make small bursts of air from the
syringe.  The resulting necks have no semblance of cylindrical symmetry, but
are broadened in the same direction as the nozzle opening.
Figure~\ref{fig:slot} shows three representative outcomes. At lower burst
pressures, the neck is flattened and ``tears'' from its edges, resulting in a
scalloped appearance that reflects a history of discrete events. At higher
pressures, the neck becomes ribbon-like near pinch-off, thinning sufficiently
to initiate coalescence near its center. 

The pinch-off of air in water shows a radically new behavior: any
cylindrical asymmetry is preserved throughout the breakup process. Others have
examined the effect of initial conditions on the evolution of the neck in a
coflowing air-water jet \cite{gordillo05}, and for very large bubbles and found
that the scaling exponent $\alpha_h$ is non-universal
\cite{bergmann06}. Other singularities \cite{eggers94} have been shown to be
sensitive to noise \cite{shi94}. However, experiment and theory up to now have
ignored cylindrical asymmetry
\cite{burton05,gordillo05,leppinen05,thoroddsen05,longuet-higgins91,oguz93},
or could not show it to be a generic feature of breakup
\cite{bergmann06}. We have shown that tilting the nozzle by just $0.07^\circ$
detectably alters the outcome of breakup. When the initial asymmetry is
strongly exaggerated, the neck tears in two.  

Because of its sensitivity to cylindrical asymmetry, the pinch-off of air in
water demonstrates a kind of memory previously unanticipated in fluid
pinch-off.  Without surface tension, the detachment dynamics no longer converge
to a cylindrically symmetric solution. When other physical processes are
modeled as the formation of a singularity, we often assume that the
singularity-formation possesses all the symmetries allowed by the fundamental
laws describing the physical process. However, it is not known whether such
symmetries can be realized in the presence of arbitrary initial conditions.  In
the case of a detaching air bubble we see that the dynamics do not assume the
full symmetry allowed and that the generic breakup is three-dimensional.

We thank Francois Blanchette, Justin Burton, Detlef Lohse, Peter Taborek, and
Lei Xu for comments. This work was supported by NSF DMR-0352777 and MRSEC
DMR-0213745.

\end{document}